\documentclass[a4paper,fleqn, review]{cas-sc}

\usepackage{amsmath,amssymb,amsthm}
\usepackage{bm}
\usepackage{graphicx}
\usepackage{booktabs}
\usepackage{xcolor}
\usepackage{hyperref}
\usepackage{natbib}
\usepackage{amsthm}
\newtheorem{definition}{Definition}
\usepackage{comment}

\begin{document}


\let\WriteBookmarks\relax
\def\floatpagepagefraction{1}
\def\textpagefraction{.001}

\shorttitle{Setting differential privacy budgets by bounding posterior-to-posterior disclosure risks}
\shortauthors{Pathiraja and Reiter}

\title[mode=title]{Setting the Privacy Budget in Differential Privacy by Bounding Adversaries' Odds of Learning Sensitive Information}

\author[1]{Ruwimal Y. Pathiraja}[orcid = 0009-0006-4704-8927]
\cormark[1]
\ead{ruwimal.pathiraja@duke.edu}

\affiliation[1]{organization={Duke University},
            city={Durham},
            country={USA}}

\cortext[1]{Corresponding author}

\author[2]{Jerome P. Reiter}[orcid = 0000-0002-8374-3832]
\ead{jreiter@duke.edu}

\affiliation[2]{organization={Duke University},
            city={Durham},
            country={USA}}




\begin{abstract}
\mbox{Differential privacy is a mathematical definition of what it means to protect data subjects' privacy in data releases.  Differential privacy depends on a parameter $\epsilon$ known as the privacy budget.  The value of $\varepsilon$ determines the nature of the privacy guarantee, with smaller values generally offering more privacy.  However, reducing $\varepsilon$ also tends to decrease the accuracy of results protected with differentially private algorithms.  Setting a value for $\varepsilon$ that satisfactorily balances this risk/accuracy trade off is complicated in practice, and there is not a standard approach to doing so. In part this is because practitioners may struggle to understand the privacy guarantee afforded by  $\varepsilon$. We present an approach to interpreting and setting $\varepsilon$ in which (i) the practitioner establishes bounds on the posterior odds that adversaries can learn sensitive information, and (ii) the practitioner converts these bounds to values of $\varepsilon$. We illustrate the approach using data from a case control study.}
\end{abstract}


\begin{keywords} confidentiality, disclosure, ratio, risk. 
\end{keywords}

\maketitle


\section{Introduction}

When releasing results of statistical analyses to the public, data stewards often are obligated to protect the confidentiality of data subjects' identities and sensitive attributes.  One way to do so is to ensure that results satisfy the privacy criterion known as differential privacy \citep{dwork:nissim, wood2018differential}.  Intuitively, this criterion requires that any released statistic takes on similar values regardless of whether any individual is in  the data used to compute the statistic.  In this way, ill-intentioned data users, henceforth called adversaries, are not able to tell whether some individual is a member of the underlying confidential database.  

There are several variants of differential privacy (DP).  We consider the original variation known as pure DP, as described in Section \ref{sec:prelim}.  In pure DP, the data steward has to select the value of a parameter $\varepsilon$ known as the privacy budget.  Smaller values of $\varepsilon$ generally correspond to stronger privacy guarantees; however, they also tend to introduce more noise into the released statistics.  Thus, data stewards seek to select an $\varepsilon$ that provides a satisfactory balance between privacy protection and data usefulness.
In practice, however, data stewards can find it challenging to specify $\varepsilon$, as its interpretation is somewhat complicated. Thus, it can be beneficial to map privacy budgets to interpretable quantities from statistical disclosure risk assessments \citep{dewaal, hotz:manski, kifer:abowd}.
In this article, we focus on the quantity known as posterior-to-posterior  disclosure risk, i.e., the ratio of the adversary's posterior probabilities of learning 
information about a targeted individual when that individual is in or is not in the data.  See \cite{kifer:abowd} for a discussion of the potential advantages of using  posterior-to-posterior disclosure risk, which we abbreviate as P2P risk, over other statistical disclosure risk measures.

Our strategy for setting $\varepsilon$ proceeds as follows. 
We first define  
two P2P risks conditional on the data release, one for the adversary learning that a targeted individual's confidential datum is contained in some disclosure set of sensitive values and the other for the adversary learning the complementary event.
  We take the ratio of these two P2P risks, which we interpret as the posterior odds ratio of learning the sensitive information when the individual does or does not participate in the data.  Large values of the posterior odds ratio may suggest that the data release makes participation in the dataset too risky for the data subject, so that the data steward may want to alter the data release strategy (e.g., reduce the privacy budget). We show that this posterior odds ratio can be written as the product of a factor that depends on $\varepsilon$ in the DP mechanism and an assumption about the adversary's prior odds ratio for learning the sensitive information.  
As described in Section \ref{sec:theory}, we presume that the data steward establishes a function that defines the maximum tolerable increase in the posterior odds ratios for 
any prior odds ratio; we call this function an odds risk profile.  For example, for an adversary whose prior odds ratio is only 0.1, the data steward may tolerate a ten-fold (or more) increase in the posterior odds ratio. In contrast, for an adversary whose prior odds ratio is 1, the data steward may not want the posterior odds ratio to increase by more than a factor of 2. The data steward can use the relationship between ratios of P2P risks and the DP budget to find values of $\varepsilon$ that accord with the desired profile for every possible value of the adversary's prior odds ratio, and use the smallest $\varepsilon$ among them for the DP  release.  
  We illustrate this approach empirically in Section \ref{sec:examples} in the context of releasing differentially private statistics from case-control studies.
  
  Our work is related to other interpretations of the privacy budget in DP. \cite{kazan2024prior} 
  relate $\varepsilon$ to  bounds on the ratio of the adversary's 
posterior probability of disclosure over their prior probability of disclosure. They define risk profiles by asking data stewards to establish these bounds for all possible prior distributions. These risk profiles are used for setting $\varepsilon$.  Our strategy follows a similar logic; however, our profiles are based on posterior-to-posterior risks rather than posterior-to-prior risks. While neither type of risk profile is necessarily ``better'' than the other, some data stewards may prefer to work with and interpret P2P risks, as recommended by \citet{kifer:abowd}. 
\cite{kifer2022bayesian} relate $\varepsilon$ and P2P risks in what is known as the bounded DP setting. They do not consider setting $\varepsilon$ via establishing risk profiles like we do here.  
\cite{lee2011much}  use a Bayesian approach to set $\varepsilon$ in
settings where the disclosure that some  individual is in the dataset is the only concern. Our strategy covers a broader setting including the incorporation of risk profiles.  

\section{Review of Differential Privacy and Posterior-to-Posterior Disclosure Risk} \label{sec:prelim}



 Let $D$ be a database comprising $n$ individuals.  We say that $D'$ is a neighboring database of $D$ when $D' = D \cup y$,  where $y$ is one additional record's data. That is, $D$ and $D'$ are neighboring databases if they have $n$ records in common and $D'$ has one additional record. 
This definition of neighboring databases leads to what is known as unbounded DP. It is also possible to define neighboring databases where $D$ and $D'$ both comprise $n$ individuals, whereby they have $n-1$ records in common and differ in one record.  This definition of neighboring databases leads to what is known as bounded DP.  In this article, we work with unbounded DP.

Let $\mathcal{M}$ be a randomized algorithm that takes as input some database $D$ and produces some output $L$.  
We say that $\mathcal{M}$ is 
$\varepsilon$-differentially private if for all
$L \subseteq \operatorname{Range}(\mathcal{M})$ and for all neighboring databases $D$ and $D'$, we have 
\begin{equation}\label{eq:dp}
\Pr[\mathcal{M}(D) \in L]
\le
\exp(\varepsilon)\Pr[\mathcal{M}(D') \in L] .
\end{equation}
The probabilities are taken with respect to the randomness in $\mathcal{M}$. The bound in  \eqref{eq:dp} must hold for all neighboring databases and output sets.  The literature on DP recommends $\varepsilon < 1$, although in practice larger values are used. 

\cite{kazan2024prior} express the DP guarantee in terms of sampling from a finite population. Using their notation, let \textbf{P} be a population of $N$ individuals. For $i=1, \dots, N$, let $y_i$ be the study 
variables measured for individual $i$.  For ease of notation, we presume each $y_i$ is a scalar.
The data steward, henceforth referred to as the agency, has data on a sample of $n$ individuals from \textbf{P}. 
For $i=1, \dots, N$, 
let $I_i = 1$ when individual $i$ is in the sample, and let $I_i=0$ otherwise. We refer to the data in the sample as $D_{obs} = \{y_i: I_i=1; i = 1, \dots, N\}$. For all $i$ with $I_i = 1$, let $D_{-i}$ 
include the study variables for all records except record $i$, that is, $D_{-i} = D_{obs} \setminus y_i$. 

The agency seeks to compute some statistic $T(D_{obs})$ and release a DP version $T^*(D_{obs})$ of it.  Let $Y_i$ and $Y_{-i}$ represent possible values that individual $i$ and the remaining sampled units could take on, respectively; let $Y = (Y_i, Y_{-i})$ represent the implied possible sample.  Writing $D = Y, D' = Y_{-i}$, $L = \{t^*\}$, and $T^*$ as the randomized algorithm $\mathcal{M}$, we can define DP in \eqref{eq:dp} as follows.   Over all values of $t^*$, all values of $Y_i = y$, 
and all values of $Y_{-i} = y_{-i}$,
we have  
\begin{equation}\label{eq:dp_unbounded}
\exp(-\varepsilon)
\le
\frac{
P\!\left[T^{*}(Y)=t^{*}\mid Y_i=y,\; I_i=1,\; Y_{-i}=y_{-i}\right]
}{
P\!\left[T^{*}(Y_{-i})=t^{*}\mid I_i=0,\; Y_{-i}=y_{-i}\right]
}
\le
\exp(\varepsilon).
\end{equation}
The two sides of the inequality follow since $Y$ and $Y_{-i}$ are interchangeable with respect to the definition of DP.

To ensure the DP guarantee, a general strategy is to construct $T^*(D_{obs})$ by adding specifically calibrated noise to $T(D_{obs})$. One example of this approach is 
the Laplace mechanism \citep{dwork2014algorithmic}. 
We compute 
$T^*(D_{obs}) = T(D_{obs}) + \eta$, 
where $\eta$ is a random draw from the Laplace$(\Delta /\varepsilon)$ distribution. 
Here, $\Delta$ is the $L_1$-sensitivity of $T(D_{obs})$, defined as the maximum amount the statistic $T(D)$ can change, i.e., $|T(D) - T(D')|$, over all possible neighboring databases $D$ and $D'$.
For example, $\Delta = 1$ when $T(D_{obs})$ is the sum of a binary variable.



To relate the DP guarantee to P2P risk, 
a key quantity is the adversary's posterior probability of learning that the value of a sensitive $y_i$ for some targeted individual $i$ is in some set $S$ of values that the agency deems a disclosure, given what is released by the agency. Throughout, we assume that the agency releases the nature of the mechanism used to create $T^*(D_{obs})$; for example, the agency indicates that it used a Laplace mechanism with global sensitivity $\Delta$ and privacy budget $\varepsilon$. 
Suppose that the adversary has a model $M$ that they use to make inferences about the confidential data.  Here, $M$ includes any auxiliary information known by the adversary.  We denote by $P_M$ the probabilities that pertain to the adversary's model. 
To characterize $M$, we re-use notation and allow  
$Y_i$ and $I_i$ to represent the adversary's  random variables in computations of disclosure risk.  Hence, probabilities like $P_M(Y_i=y | T(D_{obs}) = t^*)$ are interpreted as statements about the  adversary's posterior beliefs about the value of the study variable for individual $i$. With these interpretations in mind, we define the posterior-to-posterior disclosure risk in Definition \ref{def:p2p}.

\begin{definition}[Posterior-to-Posterior Disclosure Risk] \label{def:p2p}



For fixed data $Y$, individual $i$, adversary's model $M$, set of information $S$ that the adversary will discover after the release, and released $T^{*}(Y) = t^{*}$, the 
posterior-to-posterior disclosure risk that $Y_i \in S$ is 
\begin{equation}\label{eq:R_postpost_def}
R_i = \frac{P_M[Y_i \in S | T^*(Y) = t^*]}{P_M[Y_i \in S | T^*(Y_{-i}) = t^*]}.
\end{equation}
\end{definition}
The P2P risk  compares the probability of the adversary learning that  $Y_i \in S$ in a world in which individual $i$ is a member of the dataset with the probability of the adversary learning that  $Y_i \in S$ in a counterfactual world in which the individual is not a member of the dataset. In this context, $T^*(Y)$ and $T^*(Y_{-i})$ 
define two counterfactual releases, one in which the mechanism is applied to a dataset including individual $i$ and one in which it is applied to a neighboring dataset excluding individual $i$. The P2P risk in \eqref{eq:R_postpost_def} does not inherently incorporate the effect of  observing $I_i = 1$ or $I_i = 0$; that 
effect is discussed in Section \ref{sec:theory}. A global bound for the P2P risk in \eqref{eq:R_postpost_def} is shown in Appendix \ref{app:globalbound}.

\section{Bounds on Posterior-to-Posterior Risk Ratios}\label{sec:theory} 



In this section, we develop a strategy for setting $\varepsilon$ in an unbounded DP setting by allowing agencies to specify odds risk profiles based on  P2P risks.  To do so, we 
first define two P2P risks, which we call $R_{1,i}$ for the case where $Y_i \in S$ and $R_{0,i}$ for the case where $Y_i \notin S$.   For notational convenience, we write $T^*$ in place of $T^*(Y)$.  We define 
\begin{eqnarray}
R_{1,i} = \frac{P_M[Y_i \in S | T^* = t^*, I_i = 1]}{P_M[Y_i \in S|T^* = t^*, I_i =0]}, \hspace{10pt}
R_{0,i} = \frac{P_M[Y_i \notin S | T^* = t^*, I_i = 1]}{P_M[Y_i \notin S|T^* = t^*, I_i =0]}.
\end{eqnarray}
The key quantity in our approach is their ratio, $R_{10,i} = R_{1,i}/R_{0,i}$.  This quantity can be re-expressed as an odds ratio, since
\begin{eqnarray}
R_{10,i} &=& \left(\frac{P_M[Y_i \in S | T^* = t^*, I_i = 1]}{P_M[Y_i \in S|T^* = t^*, I_i =0]}\right)\left(\frac{P_M[Y_i \notin S | T^* = t^*, I_i = 1]}{P_M[Y_i \notin S|T^* = t^*, I_i =0]}\right)^{-1} \\
&=& \left(\frac{P_M[Y_i \in S | T^* = t^*, I_i = 1]}{P_M[Y_i \notin S|T^* = t^*, I_i =1]}\right)\left(\frac{P_M[Y_i \in S | T^* = t^*, I_i = 0]}{P_M[Y_i \notin S|T^* = t^*, I_i =0]}\right)^{-1} \label{eq:oddsratio}
\end{eqnarray}
The odds ratio in \eqref{eq:oddsratio} offers useful interpretations.  When $R_{10,i}$ exceeds one, e.g., $R_{10,i}>3$,  the adversary has greater odds of learning the individual's sensitive outcome $Y_i \in S$ when individual $i$ is in the dataset than when individual $i$ is not in the dataset. 
When $R_{10,i}$ is near or less than one, the adversary's odds are not much improved  if the individual participates in the dataset. This latter case is desirable from the perspective of privacy protection. 

We now show that $R_{10,i}$ is intimately related to the DP privacy guarantee.
To do so, we make the simplifying assumption that, for any $y_{-i}$ and $y$, 
\begin{equation}\label{eq:assump_Yminus_i}
P_M[Y_{-i} = y_{-i} |I_i = 1, Y_i = y] = P_M[Y_{-i} = y_{-i} |I_i = 0].
\end{equation}
The assumption in \eqref{eq:assump_Yminus_i}   captures the idea that the adversary's beliefs about the rest of the individuals in the dataset do not depend on the presence or absence of a particular individual or on that individual's value. The suitability of this assumption depends on the context. For example, in a large dataset in which the confidential information about each individual is their cancer status, knowing the cancer status or the presence of a particular individual in the dataset could minimally affect the adversary's beliefs about the distribution of the other individuals' statuses. However, in a small dataset in which the adversary may have auxiliary information about the total number of individuals with cancer, or in a dataset where observations are not independent, the knowledge of a particular individual's status could affect the distribution of $Y_{-i}$. 

We also formalize the assumption of a rational adversary. That is, under the adversary's model $M$ and for any possible datasets $y_{-i}$ and $y_{-i}\cup \{y_i\}$, we assume that 
\begin{eqnarray}\label{eq:assump_mech1}
P_M[T^*(y_{-i} \cup \{y_i\}
) = t^*] &=& P[T^*(y_{-i} \cup \{y_i\}) = t^*]\\
\label{eq:assump_mech2}
P_M[T^*(y_{-i}) = t^*] &=& P[T^*(y_{-i}) = t^*].
\end{eqnarray}
These two conditions encode adversaries whose model for how $T^*$ is generated is the actual DP mechanism used by the agency for the data release. 
This is reasonable 
since the DP mechanism is public.

Although not overt, the adversary's prior information is embedded in the expressions for $R_{10,i}$.  To see this, let $O_{1,i}$  be the 
odds of learning $Y_i \in S$ when individual $i$ is part of $Y$.  We write 
\begin{eqnarray}\notag
O_{1,i} &=& \frac{P_M[Y_i \in S | T^* = t^*, I_i = 1]}{P_M[Y_i \notin S|T^* = t^*, I_i =1]} \\
&=& \left(\frac{P_M[T^* = t^*| Y_i \in S, I_i = 1]P_M[Y_i \in S|I_i =1 ]}{P_M[T^*=t^*|I_i = 1]}\right) \left(\frac{P_M[T^* = t^*|I_i = 1]}{P_M[T^* = t^*| Y_i \notin S, I_i = 1]P_M[Y_i \notin S|I_i =1 ]}\right) \notag \\
&=& \left(\frac{P_M[T^* = t^*| Y_i \in S, I_i = 1]}{P_M[T^* = t^*| Y_i \notin S, I_i = 1]}\right) \left(\frac{q_i}{1-q_i}\right),\label{eq:O1_bayes}
\end{eqnarray}
where $q_i = P_M[Y_i \in S|I_i = 1]$ is the adversary's prior probability of learning $Y_i \in S$ when individual $i$ is in the data.




Similarly, let $O_{0,i}$ be the corresponding odds when individual $i$ is not part of $Y$, which we write as  
\begin{eqnarray}
O_{0,i} &=& \frac{P_M[Y_i \in S|T^* = t^*, I_i = 0]}{P_M[Y_i \notin S| T^* = t^*, I_i =0]} 
= \left(\frac{P_M[T^* = t^*| Y_i \in S, I_i = 0]}{P_M[T^* = t^*| Y_i \notin S, I_i = 0]}\right) \left(\frac{q_i^{(0)}}{1-q_i^{(0)}}\right),
\end{eqnarray}
where $q_i^{(0)} = P_M [Y_i \in S|I_i = 0]$ is the adversary's prior probability of learning $Y_i \in S$ even though individual $i$ is not in the data. Per \eqref{eq:assump_mech2}, we presume that $P_M[T^* = t^*| Y_i \in S, I_i = 0] = P_M[T^* = t^*| Y_i \notin S, I_i = 0]$, since the value of $Y_i$ should not affect the  released statistic when individual $i$ is  not in the data.  Thus, we can write $R_{10,i}$ as an odds ratio,
\begin{equation}\label{eq:R_odds_def}
R_{10,i} = \frac{O_{1,i}}{O_{0,i}} = \left(\frac{P_M[T^* = t^*| Y_i \in S, I_i = 1]}{P_M[T^* = t^*| Y_i \notin S, I_i = 1]}\right)  \left[\left(\frac{q_i}{1-q_i}\right)  \left(\frac{1-q_i^{(0)}}{q_i^{(0)}}\right)\right].
\end{equation}

Per \eqref{eq:assump_mech1}, we assume that the adversary 
uses the probability distribution for the agency's DP mechanism to compute $P_M[T^* = t^*| Y_i \in S, I_i = 1]$ and $P_M[T^* = t^*| Y_i \notin S, I_i = 1]$.    Thus, the first term in \eqref{eq:R_odds_def} 
is a function of $\varepsilon$.  The second term in \eqref{eq:R_odds_def} is the adversary's prior odds ratio, $P_{10,i} = \left(q_i/(1-q_i)\right)  \left((1-q_i^{(0)})/q_i^{(0)}\right)$.  Before discussing this quantity, we note that the index $i$ allows the possibility of individual-specific odds. In practice, however, we expect  agencies to treat individuals as exchangeable for purposes of setting $\varepsilon$.  Therefore,  in most of the following exposition, we drop the subscript $i$ to simplify notation.
Returning  to the interpretation of $P_{10}$, a value of $P_{10}=1$ indicates that the adversary's prior odds that $Y_i \in S$ are the same regardless of whether or not individual $i$ is part of the data. This is a sensible choice for $P_{10}$ when participation in the dataset is unrelated to the study variable, for example, it is completely random. 
Values of $P_{10} > 1$ (alternatively, $P_{10}<1$) model an 
adversary who is more (less) likely to uncover an individual's confidential information $Y_i \in S$ if they know the individual is a study participant than if they know otherwise.  This could arise, for example, in informative sampling schemes.  With these interpretations in mind, we can view $R_{10}$ as the adversary's updated odds ratio over $P_{10}$ after seeing the released statistic $T^*$.  

The agency may want to bound how much releasing $T^*$ increases (decreases) that posterior odds ratio.  In practice, of course, the agency generally does not know the adversary's $P_{10}$; it depends on the adversary's prior information.  This issue motivates our construction of risk profiles for $R_{10}$ for selecting $\varepsilon$, as we now describe.


Any value of $R_{10}$ can correspond to many values of $O_1$ and $O_0$, as long as their ratio is constant. However, the agency may not demand the same value of $R_{10}$ for every  possible $P_{10}$.  
For example, for an adversary who has a prior odds ratio of $P_{10}=1$, i.e., they put the same odds on $Y_i \in S$ regardless of whether or not the individual $i$ participates in the data, the agency may not want the release of $T^*$ to increase the posterior odds ratio by much, say more than a factor of two or three. The agency could view such an increase as too large, i.e., $T^*$ could enable too much refinement of the adversary's odds, for an individual who is in fact in the data.  In contrast, for an adversary who already has a very large prior odds ratio, e.g.,  $P_{10} = 10$, the agency may not consider increases in the odds ratio by some finite amount, say a factor of five, as particularly problematic.  This adversary already has strong prior beliefs so that seeing $T^*$ would not practically change their opinions about the odds ratio. 
Finally, for an adversary who has a prior odds ratio of only $P_{10} = 0.1$---this models an adversary who (perhaps oddly) believes a priori that an individual who participates in the dataset is ten times less likely to have $Y_i \in S$ than an individual who does not participate in the dataset---the agency may tolerate up to a ten-fold increase in  
$R_{10}$, as the posterior odds ratio still would be no larger than one. In other words, the agency may tolerate the additional risk from releasing $T^*$ if, 
in the end, the adversary's odds of $Y_i \in S$ remain smaller when individual $i$ participates in the data than when they do not. The agency may deem this risk acceptable.  

Recognizing this, we suppose the agency specifies a bound on the increase in $R_{10}$ for any $P_{10}$, which we write as $B(P_{10})$.  The agency's $B(P_{10})$ is a function of $P_{10}$ such that the release satisfies $R_{10} \leq B(P_{10})P_{10}$. We generally set $B(P_{10}) > 1$, possibly allowing it to get arbitrarily large but finite for some $P_{10}$.  For example, as discussed previously, the agency may consider an adversary with $P_{10} \geq 10$ already to have strong prior knowledge about the odds ratio and is not bothered if the data release increases the odds ratio by a factor even up to  $B(P_{10})=5$.  We call the collection of bounds $\mathcal{B} = \{B(P_{10})\}$ over all values of $P_{10}$ the agency's odds  risk profile. 

\begin{figure}[pos=htbp]
\begin{center}
  \includegraphics[width=0.85\textwidth]{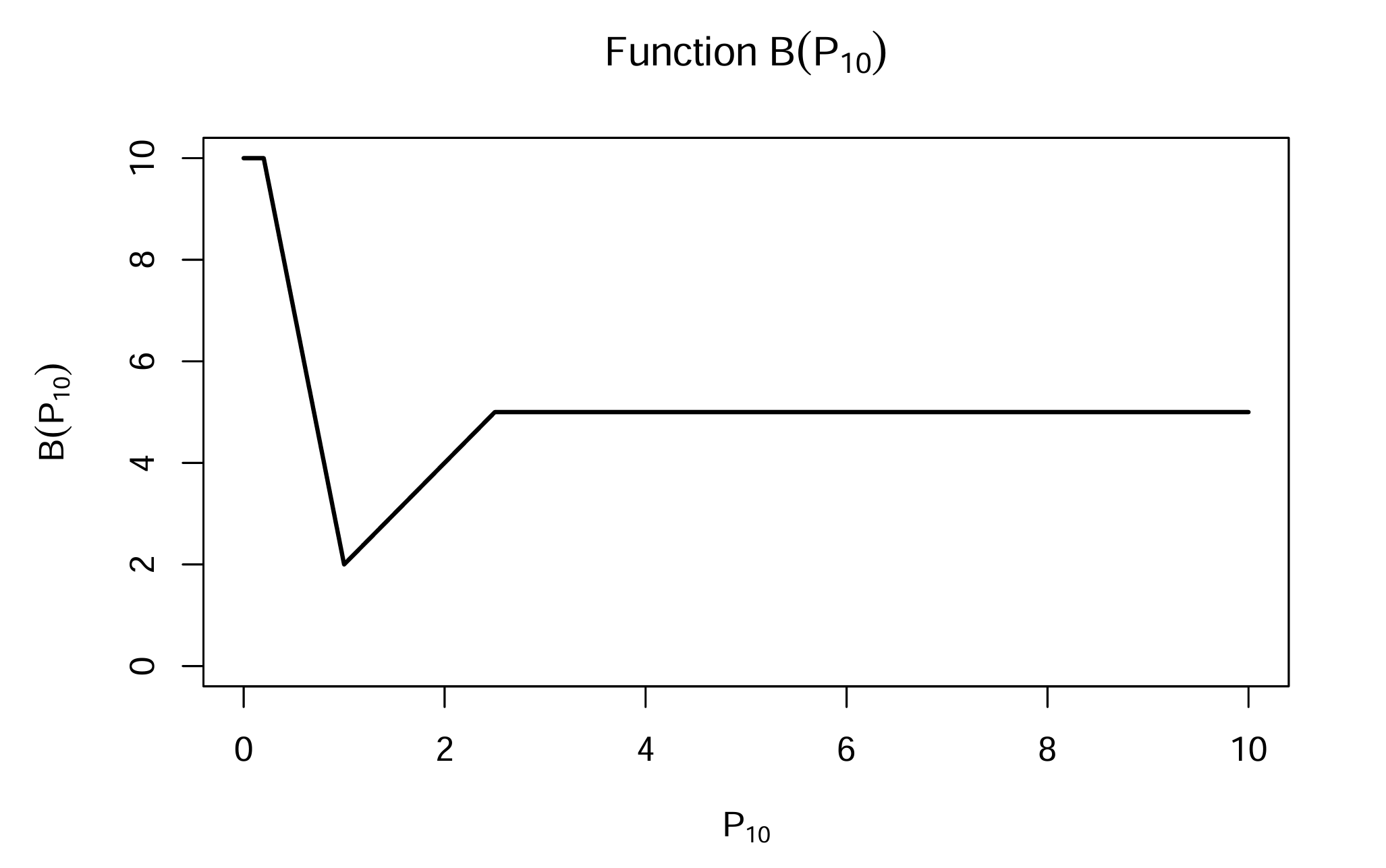}
\end{center}
\caption{Example of an agency's risk tolerance profile for increases in the odds ratio.} \label{fig:profile}
\end{figure}

Conceptually, $\mathcal{B}$ could be represented by any function.  Practically, we expect agencies to prefer simple and explainable functions. 
As an example, 
the agency might accept the bounds displayed in Figure \ref{fig:profile} and expressed as
\begin{equation}\label{eq:Rmax_piecewise}
B(P_{10}) =
\begin{cases}
\min\!\left(10,\,12-10P_{10}\right), & P_{10} \le 1,\\[4pt]
\min\!\left(5,\,2P_{10}\right), & P_{10} > 1.
\end{cases}
\end{equation}
With this $\mathcal{B}$, the agency tolerates  at most a ten-fold increase in the odds ratio caused by releasing $T^*$ for any adversaries with very low prior odds ratios.  This tolerance decreases linearly until the prior odds ratio of $P_{10}=1$, at which point the agency tolerates at most a doubling of the adversary's prior odds.  For adversaries whose prior odds ratio exceeds one, the agency is willing to tolerate doubling  the odds ratio from releasing $T^*$ up to a maximum  five-fold increase.  Of course, agencies may have other preferences; this is only one illustrative example.


The bounds $\mathcal{B}(P_{10})$ also can be interpreted as bounds on the odds risk to the individual for participating in $D_{obs}$ at any given $q_i$ and $q_i^{(0)}$ consistent with $P_{10}$. Any selected $\varepsilon$ corresponds to a bound on $O_{1}$ for a given $O_0$; that is, $\varepsilon$ governs how much the odds risk for the individual  increases by participating in $D_{obs}$ versus not participating in $D_{obs}$. This interpretation can be  useful when the agency is able to specify  meaningful values of $(q_i, q_i^0)$ based on aspects of the data release setting at hand.  We illustrate this interpretation in the context of a case-control study in Section \ref{sec:examples}.
The agency can use the relationship between $R_{10}$ and $\varepsilon$ to find the privacy budget that accords with the desired profile. 
The agency uses the DP mechanism and \eqref{eq:R_odds_def} to solve for $\varepsilon$ for every possible value of $P_{10}$ and corresponding  $\mathcal{B}(P_{10})$ from the risk profile.  The agency  uses the smallest value of $\varepsilon$ among these solutions to release $T^*(D_{obs})$. When using odds risk profiles specific to individuals, the agency uses the smallest value of $\varepsilon$ across the profiles. 

We now show how $\varepsilon$ is computed in terms of the parameters above.
To do so, we make use of Lemma 1 in \cite{kazan2024prior}, which we paraphrase here for completeness.  Under  the 
assumptions in \eqref{eq:assump_Yminus_i}--\eqref{eq:assump_mech2}, if the release of $T^{*}=t^{*}$ satisfies $\varepsilon$-DP, then for any subset $S$ of the domain of $Y_i$, we have 
\begin{align}
\exp(-\varepsilon)
&\le
\frac{
P_{\mathcal{M}}\!\left[T^{*}=t^{*}\mid Y_i \in S,\; I_i=1\right]
}{
P_{\mathcal{M}}\!\left[T^{*}=t^{*}\mid I_i=0\right]
}
\le
\exp(\varepsilon),
\label{eq:kazan_lemma_one}
\\[6pt]
\exp(-2\varepsilon)
&\le
\frac{
P_{\mathcal{M}}\!\left[T^{*}=t^{*}\mid Y_i \in S,\; I_i=1\right]
}{
P_{\mathcal{M}}\!\left[T^{*}=t^{*}\mid Y_i \notin S,\; I_i=1\right]
}
\le
\exp(2\varepsilon).
\label{eq:kazan_lemma_two}
\end{align}
From  \eqref{eq:kazan_lemma_one} and \eqref{eq:kazan_lemma_two}, and dropping the subscript for unit $i$, we have 
$\exp(-2\varepsilon) P_{10} \leq R_{10} \leq \exp(2\varepsilon) P_{10}$.
For any $P_{10}$, 
we  find 
the largest value of $\varepsilon$, 
say $\varepsilon_{P_{10}}$, that accords with 
this inequality. Since the risk profile  requires $R_{10} \leq B(P_{10})P_{10}$, this largest value is found as 
\begin{equation}\label{eq:cond_B}
\exp(2\varepsilon_{P_{10}})  P_{10} = B(P_{10}) P_{10} \implies  \varepsilon_{P_{10}} = (1/2)   \log\left(B(P_{10}) \right).
\end{equation}
For \eqref{eq:cond_B} to hold, we require $1 \leq B(P_{10})$. Thus, the risk profiles should adhere to this condition.  

The result in \eqref{eq:cond_B} applies for a specific $P_{10}$. The agency may desire a privacy budget that adheres to the risk profile for all values of $P_{10}$. 
To determine this privacy budget, the agency searches over $0< P_{10} < \infty$ to find $\varepsilon_{min} = \min_{P_{10}} \varepsilon_{P_{10}}$. The agency then uses $\varepsilon_{min}$ in the data release.





These formulas apply for any $\varepsilon$-DP algorithm and any function of $D_{obs}$. When $T^*(D_{obs})$ is multivariate, the agency could use the selected $\varepsilon_{min}$ as the total privacy budget for the data release. 

\section{Application to Case Control Studies}\label{sec:examples}


In this section, we illustrate how agencies can use the bounds on posterior odds ratios in the context of a case-control study. Suppose the agency designs a study comprising $2n+1$ individuals. To do so, the agency initially samples $n$ cases (i.e., individuals with $y_i=1$) and $n$ controls (i.e., individuals with $y_i=0$) from the population.  The agency randomly chooses the final individual in the study to be either a case or control. The agency reveals the value of $2n+1$ but does not reveal the numbers of cases and controls ultimately sampled.  In this way, if an adversary somehow knew the disease status for all but one individual in the study, that adversary would not be able to tell with certainty from the information released about the study design whether this final individual has the disease or not.  We presume that the $2n+1$ individuals are sampled from a population comprising many more than $2n+1$ cases and controls.

 We first consider how to set $\varepsilon$ using the methods in Section \ref{sec:theory} for a specific set of 
  adversary's prior odds. 
 According to the sampling design, a rational  adversary should set $q_i = P_M[Y_i =1|I_i = 1] = 0.5$ for any unit $i$ in the population. As a reasonable value for $q_i^{(0)} = P_M[Y_i=1|I_i = 0]$, the agency can presume an adversary who sets it equal to the prevalence of the disease in the (large) population, which we denote by $\pi$.  Thus, for this adversary, we have $P_{10} = (1 - \pi)/\pi$.  As another example, the agency instead may model adversaries who know nothing about the prevalence of the disease and naively presume $q_i^{(0)} = 0.5$ for all individuals in the population.  For this adversary, $P_{10} = 1$.  Finally, the agency can consider adversaries with arbitrary values of $q_i^{(0)}$ and hence arbitrary values of $P_{10}$, with the goal of ensuring the selected $\varepsilon_{min}$ satisfies a risk profile for any adversary.
 

We now illustrate these computations. To use realistic numbers, we use information from the case-control study of \cite{duerr2006genome}. This study aims to identify genetic factors that might contribute to Crohn's Disease.  The study uses a total sample size of 1095 individuals, about half of whom have the disease and half do not.  Thus, for our illustration, we set $2n+1 = 1095$. 
Using their data, we approximate the population prevalence of Crohn's Disease 
among the population 
considered by  \cite{duerr2006genome} 
as $\pi = 0.00125$.  
Thus, to model an adversary who bases $q_i^{(0)}$ on the prevalence of the disease, 
we set 
$q_i^{(0)}
\approx 0.00125$ and $P_{10} \approx 
800$ for our illustration.

To demonstrate how to turn values of $P_{10}$ into choices for $\varepsilon$, we suppose that the agency uses the risk profile $\mathcal{B}$  defined in \eqref{eq:Rmax_piecewise}. For the adversary who uses the prevalence of the disease, this profile sets a bound of $B(800) = 5$. 
Solving \eqref{eq:cond_B}, the agency should ensure $\varepsilon_{800} \leq (1/2) \log(5) \approx 0.805$  to satisfy the risk profile for this adversary. 
For the adversary who sets $P_{10}=1$, the risk profile sets the bound to $B(1) = 2$. To satisfy the risk profile for this adversary, the agency should set $\varepsilon_1 = (1/2) \log(2) \approx 0.347$.  

\begin{figure}[pos=htbp]
\begin{center}
  \includegraphics[width=0.85\textwidth]{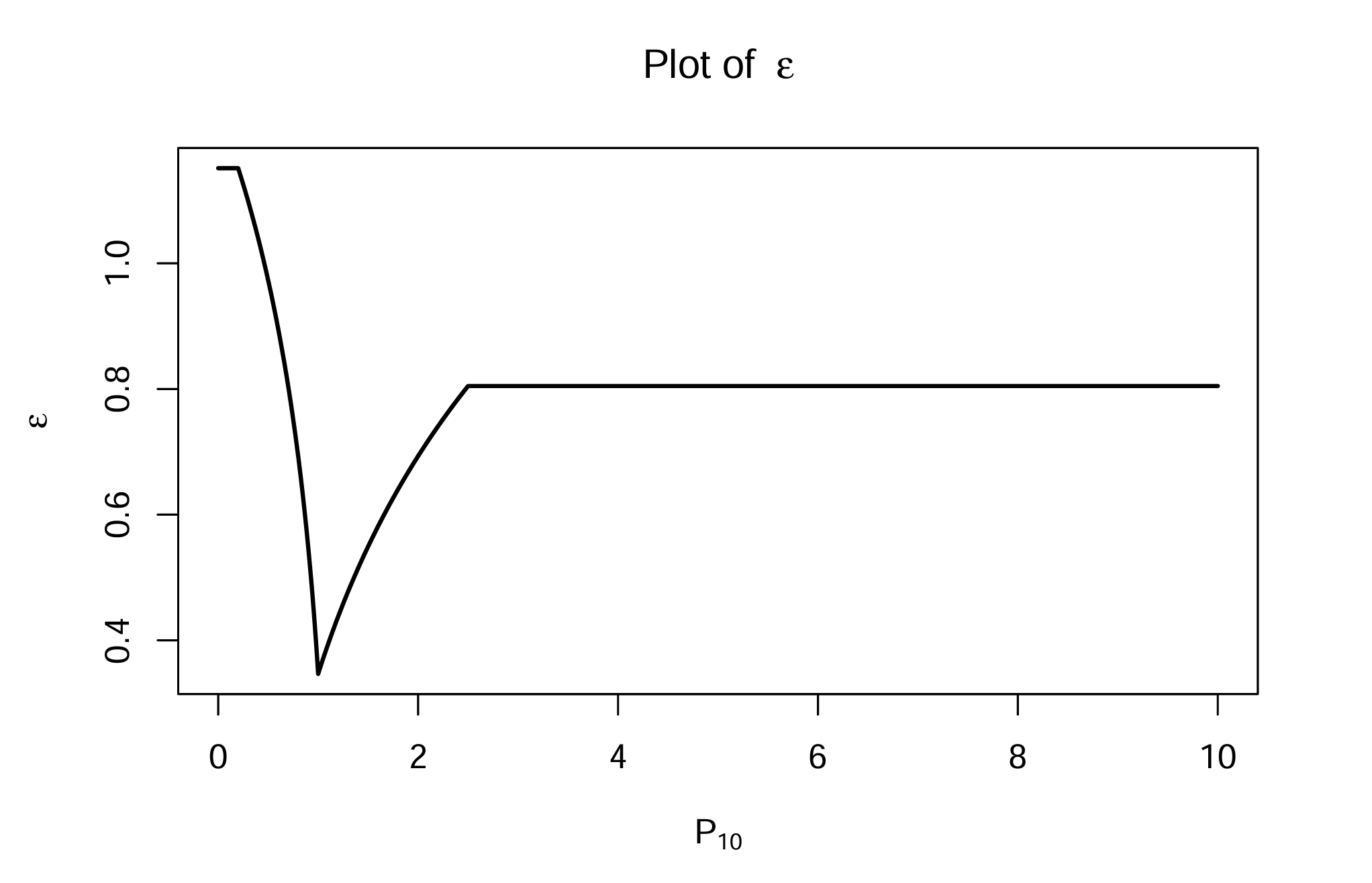}
\end{center}
\caption{Implied values of $\varepsilon$ for different adversary's prior odds ratios using the illustrative risk profile. The agency would use the smallest value of $\varepsilon$ for the data release, which occurs when the prior odds $P_{10}=1$ for this profile.}\label{fig:epsilon}
\end{figure}

These determinations of $\varepsilon$ are based on two specific assumptions about $P_{10}$. 
The agency may want to use a privacy budget that satisfies the risk profile for any adversary's $P_{10}$.   Figure \ref{fig:epsilon} displays the implied value of $\varepsilon_{P_{10}}$ for each $P_{10}$. Visually, we see that the minimum $\varepsilon_{P_{10}}$ is reached at $P_{10}=1$.  Algebraically, 
by \eqref{eq:cond_B}, the agency should select the $\varepsilon$ that satisfies $\varepsilon \leq (1/2) \log (B(P_{10}))$ for all $P_{10}$. Since $\log$ is an increasing function on $(0, \infty)$ and the profile in \eqref{eq:Rmax_piecewise} has $\min B(P_{10}) = 2$ at $P_{10} = 1$, we get $\varepsilon_{min}  
\approx 0.347$. 
The agency would use this value of $\varepsilon$, possibly after rounding it to 0.35 or 0.30 for convenience, for the data release.  Of course, with different risk profiles, the agency may require a different value of $\varepsilon$ to satisfy the implied bounds.

\section{Concluding Remarks}

We present a strategy for interpreting and setting $\varepsilon$ in differential privacy that enables the agency to incorporate the adversary's potential prior knowledge expressed in terms of odds ratios.  By searching over all possible values of the prior odds ratio, the agency can select an $\varepsilon$ that is not specific to one set of assumptions about the adversary's prior knowledge. The value of $R_{10}$ can be interpreted as
the multiplicative increase in the posterior  odds of an individual's record being compromised by participating in the dataset (study) rather than not participating in it. For example, if $R_{10} = 3$, individuals' odds ratios increase threefold by participating in the dataset as opposed to staying out of it. Using this framework may help an agency  explain the DP privacy guarantee to potential study participants.
In terms of future research, one direction  is to apply the framework 
to other definitions of DP such as R\'enyi DP. Such extensions would require converting divergence-based privacy guarantees into bounds on odds ratios. 

\section*{Funding}

This research did not receive any specific grant from funding agencies in the public, commercial, or not-for-profit sectors.

\section*{Declaration of generative AI and AI-assisted technologies in the manuscript preparation process}
 During the preparation of this work the authors used ChatGPT in order to assist with phrasing improvements, reference searching, and \LaTeX{} formatting. After using this tool/service, the authors reviewed and edited the content as needed and take full responsibility for the content of the published article.

\bibliographystyle{cas-model2-names}
\bibliography{references}




\appendix

\makeatletter
\@addtoreset{equation}{section}
\@addtoreset{figure}{section}
\@addtoreset{table}{section}
\makeatother

\renewcommand{\theequation}{\thesection.\arabic{equation}}
\renewcommand{\thefigure}{\thesection.\arabic{figure}}
\renewcommand{\thetable}{\thesection.\arabic{table}}

\section{Global Bound on Posterior-to-Posterior Disclosure Risk}
\label{app:globalbound}

In this appendix, we derive a global bound for the posterior-to-posterior disclosure risk ratio $R_i$ in \eqref{eq:R_postpost_def} 
 when the sensitive set $S \subseteq \{y_i\}$. The derivation of this global bound is similar to the derivation in \cite{kifer2022bayesian} with the distinction that we presume 
an unbounded-DP setting.  
For the numerator of $R_i$, we have  
\begin{equation}\label{eq:post_TY}
P_M[Y_i = y_i | T^*(Y) = t^*] = 
\frac{\displaystyle \sum_{y_{-i} \in \mathbf{Y_{-i}}} P_M[Y_{-i} = y_{-i}]  P_M[y_i|Y_{-i} = y_{-i}]  P[T^*(y_{-i} \cup \{y_i\}) = t^*]}{\displaystyle \sum_{y_{-i} \in \mathbf{Y_{-i}}} \sum_{y_i'} P_M[Y_{-i} = y_{-i}]   P_M[y_i'|Y_{-i} = y_{-i}]  P[T^*(y_{-i} \cup \{y_i'\}) = t^*]}.
\end{equation} 
For the denominator of $R_i$, we have 
\begin{equation}\label{eq:post_TYminus}
\begin{aligned}
P_M[Y_i = y_i | T^*(Y_{-i}) = t^*] &= \frac{\displaystyle \sum_{y_{-i} \in \mathbf{Y_{-i}}} P_M[Y_{-i} = y_{-i}]  P_M[y_i|Y_{-i} = y_{-i}]  P[T^*(y_{-i}) = t^*]}{\displaystyle \sum_{y_{-i} \in \mathbf{Y_{-i}}} \sum_{y_i'} P_M[Y_{-i} = y_{-i}]   P_M[y_i'|Y_{-i} = y_{-i}]  P[T^*(y_{-i}) = t^*]} \\
&= \frac{\displaystyle \sum_{y_{-i} \in \mathbf{Y_{-i}}} P_M[Y_{-i} = y_{-i}]   P_M[y_i|Y_{-i} = y_{-i}]  P[T^*(y_{-i}) = t^*]}{\displaystyle \sum_{y_{-i} \in \mathbf{Y_{-i}}} P_M[Y_{-i} = y_{-i}] P[T^*(y_{-i}) = t^*]}.
\end{aligned}
\end{equation}
Using \eqref{eq:post_TY} and \eqref{eq:post_TYminus}, we can write 
    $R = 
    A  B$ where
\begin{equation}\label{eq:A_def}
A = \frac{\displaystyle \sum_{y_{-i} \in \mathbf{Y_{-i}}} P_M[Y_{-i} = y_{-i}]   P_M[y_i|Y_{-i} = y_{-i}]  P[T^*(y_{-i} \cup \{y_i\}) = t^*]}{\displaystyle \sum_{y_{-i} \in \mathbf{Y_{-i}}} P_M[Y_{-i} = y_{-i}]   P_M[y_i|Y_{-i} = y_{-i}]  P[T^*(y_{-i}) = t^*]}
\end{equation}
and
\begin{equation}\label{eq:B_def}
B = \frac{\displaystyle \sum_{y_{-i} \in \mathbf{Y_{-i}}} P_M[Y_{-i} = y_{-i}]   P[T^*(y_{-i}) = t^*]}{\displaystyle \sum_{y_{-i} \in \mathbf{Y_{-i}}} \sum_{y_i'} P_M[Y_{-i} = y_{-i}]   P_M[y_i'|Y_{-i} = y_{-i}]  P[T^*(y_{-i} \cup \{y_i'\}) = t^*]}.
\end{equation}

By the guarantee of $\varepsilon$-DP  from \eqref{eq:dp_unbounded}, we have $P[T^*(y_{-i} \cup \{y_i\}) = t^*] \leq \exp(\varepsilon) P[T^*(y_{-i}) = t^*]$ and $\exp(-\varepsilon)P[T^*(y_{-i}) = t^*] \leq P[T^*(y_{-i} \cup \{y_i'\}) = t^*].$ Substituting into $A$ and $B$, respectively, we get $A \leq \exp(\varepsilon)$ and $B \leq \exp(\varepsilon)$. Therefore, $R_i\leq \exp(2 \varepsilon)$ is the upper bound for the posterior-to-posterior risk ratio. Thus,
we have $\exp(-2\varepsilon) \leq R_i \leq \exp(2\varepsilon).$

This bound relies on \eqref{eq:assump_mech1} and \eqref{eq:assump_mech2}; it does not require \eqref{eq:assump_Yminus_i} to hold. The resulting bound applies across a broader range of settings, albeit with the restriction that it is a global bound and does not take the agency's risk profile 
into account.

\end{document}